\documentclass[prb,showpacs,preprintnumbers,amsmath,amssymb,twocolumn,superscriptaddress]{revtex4-1}
\usepackage{graphicx}
\usepackage{dcolumn}
\usepackage{bm}
\usepackage{amssymb}
\usepackage{amsmath}
\usepackage{epstopdf}
\usepackage{booktabs}
\usepackage{threeparttable}
\usepackage[pdfstartview=FitH, CJKbookmarks=true, bookmarksnumbered=true, bookmarksopen=true, colorlinks, pdfborder=001, linkcolor=blue, anchorcolor=blue, citecolor=blue]{hyperref}
\begin{document}
\title{Antiferromagnetism with divalent Eu in EuNi$_5$As$_3$}
\author{W. B. Jiang}
\affiliation{Center for Correlated Matter and Department of Physics, Zhejiang University, Hangzhou, 310058, China}
\author{M. Smidman}
\email{msmidman@zju.edu.cn}
\affiliation{Center for Correlated Matter and Department of Physics, Zhejiang University, Hangzhou, 310058, China}
\author{W. Xie}
\affiliation{Center for Correlated Matter and Department of Physics, Zhejiang University, Hangzhou, 310058, China}
\author{J. Y. Liu}
\affiliation{Department of Chemistry, Zhejiang University, Hangzhou, 310058, China}
\author{J. M. Lee}
\affiliation{National Synchrotron Radiation Research Center, Hsinchu, 30076, Taiwan}
\author{J. M. Chen}
\affiliation{National Synchrotron Radiation Research Center, Hsinchu, 30076, Taiwan}
\author{S. C. Ho}
\affiliation{National Synchrotron Radiation Research Center, Hsinchu, 30076, Taiwan}
\author{H. Ishii}
\affiliation{National Synchrotron Radiation Research Center, Hsinchu, 30076, Taiwan}
\author{K. D. Tsuei}
\affiliation{National Synchrotron Radiation Research Center, Hsinchu, 30076, Taiwan}
\author{C. Y. Guo}
\affiliation{Center for Correlated Matter and Department of Physics, Zhejiang University, Hangzhou, 310058, China}
\author{Y. J. Zhang}
\affiliation{Center for Correlated Matter and Department of Physics, Zhejiang University, Hangzhou, 310058, China}
\author{Hanoh Lee}
\affiliation{Center for Correlated Matter and Department of Physics, Zhejiang University, Hangzhou, 310058, China}
\author{H. Q. Yuan}
\email{hqyuan@zju.edu.cn}
\affiliation{Center for Correlated Matter and Department of Physics, Zhejiang University, Hangzhou, 310058, China}
\affiliation{Collaborative Innovation Center of Advanced
Microstructures, Nanjing 210093, China}
\date{\today}

\begin{abstract}

   We have successfully synthesized single crystals of EuNi$_5$As$_3$ using a flux method and we present a comprehensive study of the physical properties
   using magnetic susceptibility, specific heat, electrical resistivity, thermoelectric power and x-ray absorption spectroscopy (XAS) measurements. EuNi$_5$As$_3$ undergoes two close antiferromagnetic transitions at respective temperatures of $T_{N1}$~=~7.2~K and $T_{N2}$~=~6.4~K, which are associated with the Eu$^{2+}$ moments. Both transitions are suppressed upon applying a field and we map the temperature-field phase diagrams for fields applied parallel and perpendicular to the easy $a$~axis.  XAS measurements reveal that the Eu is strongly divalent, with very little temperature dependence, indicating the localized Eu$^{2+}$ nature of EuNi$_5$As$_3$, with a lack of evidence for heavy fermion behavior.

\pacs{75.30.Kz,75.50.Ee,78.70.Dm}

\end{abstract}

\maketitle

\section{Introduction}
In recent decades, Kondo related physics has been intensively investigated in rare-earth element intermetallics.
In the Kondo lattice, the on-site Kondo interaction screens the magnetic moment of localized $f$ electrons,
leading to a nonmagnetically ordered heavy Fermion state. Besides the Kondo interaction, another competing interaction, the
Ruderman-Kittel-Kasuya-Yosida(RKKY) interaction mediated by the surrounding conduction electrons conversely favors long-range magnetic
order. The competition between the Kondo and RKKY interactions in heavy fermion systems
may result in various ground states, such as, magnetic order, superconductivity, heavy fermion and intermediate valence states \cite{Doniach,Ye,Review1,Review2,Review3}. Such phenomena related to the Kondo effect have often been observed in Ce,
Yb and U-based compounds, but Eu based heavy fermion materials have not been commonly
reported. In Eu based compounds, the Eu ion typically adopts one of two electronic configurations: divalent,
magnetic Eu$^{2+}$ (4f$^7$, J=7/2, L=0) or trivalent, nonmagnetic Eu$^{3+}$ (4f$^6$, J=0, L=3).
Therefore usually either a  magnetic state with localized Eu$^{2+}$, a non-magnetic state with valence fluctuations or trivalent Eu$^{3+}$ occur in Eu systems. However, there have been a few proposed examples of heavy fermion systems such as EuNi$_2$P$_2$
and EuCu$_2$(Si$_{1-x}$Ge$_x$)$_2$~\cite{ENP122,ECGZakir}.The Eu valence of EuNi$_2$P$_2$ also shows a significant temperature dependence with the valence changing from +2.25 at 300~K to +2.50
at 1.4~K~\cite{Nagarajan1985}, exhibiting strong valence fluctuations in the ground state. The reason for the rare existence of heavy fermion behavior in Eu based compounds remains an open question, requiring further experimental and theoretical investigations.

In this paper, we report the successful synthesis of single crystals of EuNi$_5$As$_3$
using a self flux method and study the physical properties by means of electrical
resistivity, magnetization, specific heat and partial fluorescence yield X-ray
absorption spectroscopy(PFY-XAS) measurements. Polycrystalline samples of
EuNi$_5$As$_3$ were also obtained to verify the crystal structure and to
measure the thermoelectric power $S(T)$. Our results provide evidence
for two antiferromagnetic (AFM)  transitions at $T_{N1}~=~7.2$~K and $T_{N2}~=~6.4$~K.
 The two AFM transitions are suppressed by applied  magnetic fields and we map the field-temperature phase diagrams for $H\parallel a$ and $H\perp a$. A divalent Eu valence is deduced from PFY-XAS measurements and the weak temperature
dependence of the Eu valence configuration confirms the Eu$^{2+}$ AFM ordering in EuNi$_5$As$_3$.

\section{Experimental details}

Single crystals of EuNi$_5$As$_3$ were synthesized using a NiAs self-flux method.
EuAs and NiAs were first synthesized as described elsewhere~\cite{WBJiang2015}.
Subsequently, EuAs, NiAs and Ni were combined in the ratio EuNi$_5$As$_3$:NiAs of 1:3.
The mixtures were then combined and sealed in an evacuated quartz ampoule. The ampoule was slowly
heated to 1000$^\circ$C, and held at this temperature for 24 hours to allow for homogenization
before being slowly cooled to 800$^\circ$C at a rate of 3$^\circ$C/hour, and then quickly cooled
to room temperature. Rectangular rod-like single crystals with a typical length of 2mm were
obtained after mechanical separation from the remaining flux. Single crystals of SrNi$_5$As$_3$,
which were used as a non-magnetic analog were obtained using the same procedure.
The as-grown single crystals are stable in air. To further clarify the
crystal structure, we also synthesized polycrystalline samples of EuNi$_5$As$_3$
using a solid state reaction method. EuAs, NiAs and Ni powder were combined stoichiometrically
and sintered at 800$^\circ$C for 4 days. The resulting pellet was then thoroughly ground and
pressed before being annealed at 850$^\circ$C for a week in order to improve the sample homogeneity.

The crystal structure was examined using single crystal x-ray diffraction on an Xcalibur, Atlas, Gemini
ultra diffractometer with an x-ray wavelength of $\lambda=0.71073{\AA}$. Room temperature powder x-ray
diffraction (XRD) data were collected using a PANalytical X'Pert MRD diffractometer with Cu K$_{\alpha1}$
radiation and a graphite monochromator. The chemical composition was also checked by energy-dispersive
x-ray spectroscopy using a FEI SIRION-100 field emission scanning electron microscope. Resistivity
$\rho(T,H)$, magnetization $M(T,H)$,  specific heat $C(T)$ and thermoelectric power measurements were performed using a Quantum Design Physical Property Measurement System. The Eu L$_{III}$-edge partial fluorescence yield
X-Ray absorption spectroscopy was measured at the beamline 12XU in Spring-8, Japan, as described in
Ref.~\onlinecite{WBJiang2015}

\section{Results and discussion}

\subsection{Crystal structure}

\begin{figure}[t]
     \begin{center}
     \includegraphics[width=0.9\columnwidth]{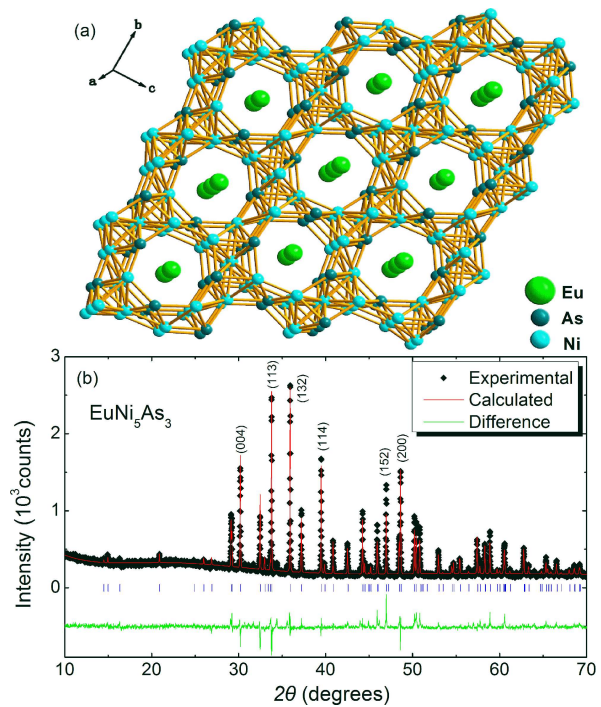}
     \end{center}
     \caption{(Color online)(a) Crystal structure of EuNi$_5$As$_3$. (b) The powder x-ray diffraction pattern of
     polycrystalline EuNi$_5$As$_3$. The solid red line shows the calculated results of the Rietveld refinement,
     while the vertical bars show the theoretical Bragg peak positions and the
     green line shows the difference between the observed and calculated data.}
     \label{fig1}
\end{figure}

Rod-like single crystals were selected for single crystal x-ray diffraction measurements at
room temperature, to clarify the crystal structure. The results of the refinement of the crystal structure
from the single crystal diffraction reflections are displayed in Table~\ref{XRDtable}. EuNi$_5$As$_3$ crystallizes in
the LaCo$_5$P$_3$-type orthorhombic structure with the space group $Cmcm$(No.63), isostructural to EuNi$_5$P$_3$ \cite{Badding1987,Probst1991}. The
reliability factors $R$ and $S$ obtained from the structural refinement are 8.59$\%$ and
1.2 respectively, confirming the accuracy of the refinement. The refined cell parameters
$a$, $b$, $c$ are 3.7381, 12.1977 and 11.8239${\AA}$ respectively, which are consistent
with the previous report~\cite{Probst1991}. As shown in Fig.~\ref{fig1}, along the [100]
direction, the Eu atoms form one dimensional-like chains and are separated by a Ni-As structure.
In addition, the powder XRD patterns for
polycrystalline EuNi$_5$As$_3$, as shown in Fig.~\ref{fig1}(b), can be well fitted using similar
structural parameters to those derived from the single crystal XRD refinement.

\begin{table}
\centering
\caption{Atomic coordinates and isotropic displacement parameters U$_{eq}$ from the refinement of single crystal
 XRD data.}
\begin{tabular}{*8c}
  \hline
  Atom & Site  & X & Y & Z & U$_{eq}$ \\ \hline
  As & 8f  & 0 & 0.3792 & 0.5436 & 0.01338 \\
  As & 4c  & 0.5 & 0.6146 & 0.75 & 0.01394 \\
  Eu & 4c  & 0.5 & 0.3348 & 0.75 & 0.01632 \\
  Ni & 8f  & 0 & 0.1937 & 0.5654 & 0.01645 \\
  Ni & 4a  & 0.5 & 0.5 & 0.5 & 0.01431 \\
  Ni & 8f  & 0 & 0.5492 & 0.6445 & 0.01527 \\
  \hline
\end{tabular}
\label{XRDtable}
\end{table}

\subsection{Magnetic order in  EuNi$_5$As$_3$}

\begin{figure}[t]
     \begin{center}
     \includegraphics[width=3.2in,keepaspectratio]{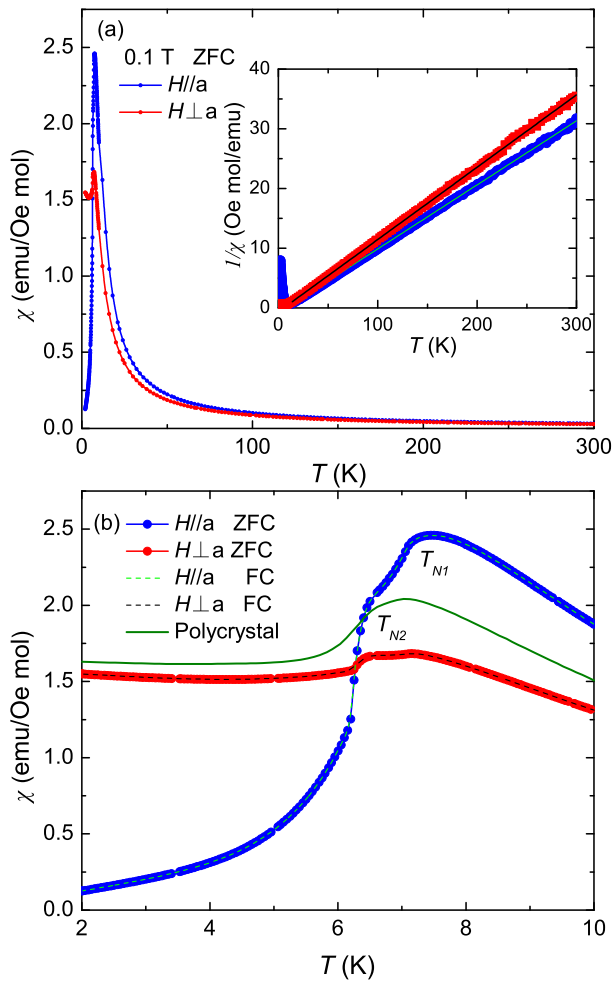}
     \end{center}
     \caption{(Color online) (a) Magnetic susceptibility as a function of  temperature of single crystals of EuNi$_5$As$_3$ with zero-field cooling.
     The inset shows $1/\chi$ against temperature and the corresponding Curie-Weiss fitting. (b) Low temperature magnetic susceptibility for single
     crystals with zero-field and field cooling, along with measurements of the polycrystalline sample.}
     \label{fig2}
\end{figure}

To characterize the magnetic properties of EuNi$_5$As$_3$, we performed magnetic susceptibility
measurements on both single crystalline and polycrystalline EuNi$_5$As$_3$. Figure.~\ref{fig2} displays the magnetic susceptibility $\chi(T)$ in an applied field of 0.1 T for $H \| a$
and $H \bot a$, where corrections for the demagnetization have been made based on the sample geometry. For both field directions, the data follows the Curie-Weiss law above 10~K.
The inverse susceptibility $1/\chi(T)$ and the fit to the Curie-Weiss law are shown in the
inset of Fig.~\ref{fig2}(a). The derived effective moments for the two field directions
are $\mu_{eff}^{a}=8.64\mu_B$ and $\mu_{eff}^{bc}=8.12\mu_B$, which are both close to the
expected 7.94$\mu_B$ for a free Eu$^{2+}$(J=7/2) ion. In addition, the obtained Curie-Weiss temperatures $\theta_P$ are -7.85~K and -5.45~K for $H \| a$ and $H \bot a$, respectively,
indicating AFM interactions between the Eu$^{2+}$ moments.
The low temperature magnetic susceptibility $\chi(T)$ is shown in Fig.~\ref{fig2}(b).
In the case of $H \| a$, $\chi(T)$ shows an AFM transition around 7.0~K with
pronounced drop at lower temperatures corresponding to another transition at 6.2~K. When $H$ is applied perpendicular to the $a$~axis, $\chi(T)$ also shows transitions at 7.0~K and 6.3~K, while at lower temperatures $\chi(T)$  remains almost constant. A decrease at the transitions when $H \| a$ and almost constant behavior when $H \perp a$ indicates that the $a$-axis corresponds to the easy direction.
The lack of hysteresis between zero-field cooling (ZFC) and field-cooling (FC) measurements
indicates that these transitions correspond to AFM ordering. The magnetic susceptibility
$\chi(T)$ of the polycrystalline sample also shows two similar transitions, supporting the single crystal results.

\begin{figure}[t]
     \begin{center}
     \includegraphics[width=3.2in,keepaspectratio]{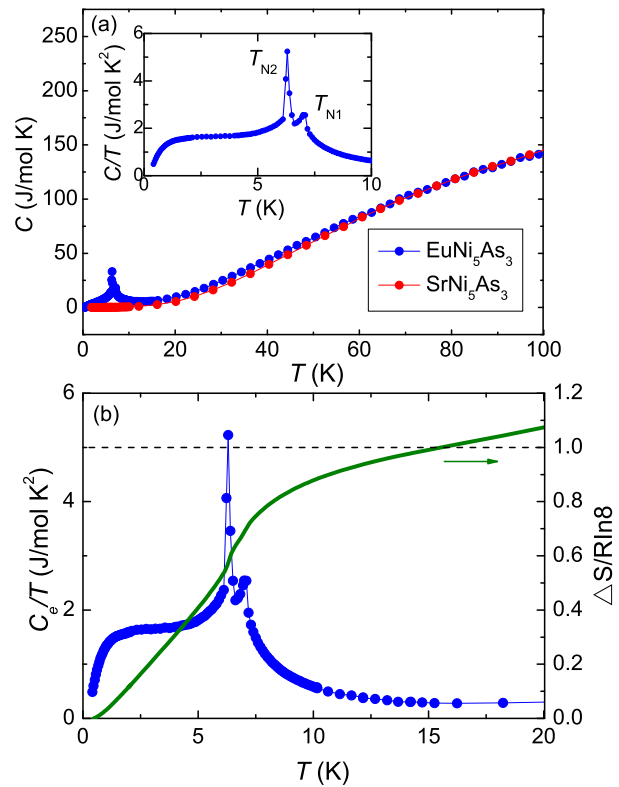}
     \end{center}
     \caption{(Color online) (a) Temperature dependence of the specific heat $C(T)$ for single crystalline
     EuNi$_5$As$_3$ and the isostructural compound SrNi$_5$As$_3$ in zero field. The inset shows the low
     temperature region displaying two AFM transitions. (b) Magnetic specific heat $C_e(T)/T$ of EuNi$_5$As$_3$  after subtracting the lattice contribution.
     The magnetic entropy $\Delta S(T)$ is also shown. The dashed line marks the position where the entropy
     is $R\ln8$. }
     \label{fig3}
\end{figure}

The main panel of Fig.~\ref{fig3} shows $C(T)$ in the temperature range 0.4~K~-~100~K for EuNi$_5$As$_3$ and the non-magnetic
isostructural compound SrNi$_5$As$_3$ for comparison. At high temperatures, $C(T)$ for both compounds overlap, indicating that the lattice contribution of EuNi$_5$As$_3$ can be taken to be the same as that
of SrNi$_5$As$_3$. It can be seen in the inset that two sharp peaks are clearly observed, which can be attributed to two AFM transitions at $T_{N1}~=~7.2$~K and $T_{N2}~=~6.4$~K, consistent with the values from $\chi(T)$. At low temperatures, $C(T)/T$
also shows a pronounced plateau around 2~K. This feature may be due to the Zeeman splitting of the $^8S_{7/2}$ multiplet
of the Eu$^{2+}$ ions in the internal magnetic field, similar to many magnetic Eu and Gd based compounds
such as, EuB$_6$, EuCu$_2$As$_2$ and Gd$_2$Fe$_3$Si$_5$~\cite{EuB6,PRBJohnston,Vining1983}.

The low temperature specific heat of non-magnetic SrNi$_5$As$_3$ was fitted (not shown) using  $C/T=\gamma+\beta T^2$. The derived electronic specific coefficient $\gamma$ of SrNi$_5$As$_3$ is 16.1~mJ/mol K$^2$, while the Debye temperature $\theta_D~=~338.4$~K was calculated using $\theta_D=\sqrt[3]{12\pi^4nR/5\beta}$, where $\beta~=~0.451~\mu$J/mol K$^4$, $n~=~9$ is the number of atoms per formula unit and $R=$~8.314~J/mol~K.  The magnetic contribution to the specific
heat of EuNi$_5$As$_3$ ($C_e/T$) is shown in Fig.~\ref{fig3}(b), which is obtained from subtracting the phonon contribution estimated
from fitting the data of SrNi$_5$As$_3$. The entropy $\Delta S(T)$ is also shown in Fig.~\ref{fig3} (b),
where $\Delta S(T)$ continuously increases but shows two anomalies around the two transitions, indicating a
second-order nature of the AFM transitions at $T_{N1}$ and $T_{N2}$. The full magnetic entropy $R\ln8$
expected for divalent Eu is recovered around 15~K, above the AFM transition temperatures, possibly due
to the formation of short range magnetic order or magnetic fluctuations, which may be seen from the upturn
of $C_e/T$ below 16~K.

\begin{figure}[t]
     \begin{center}
     \includegraphics[width=3.2in,keepaspectratio]{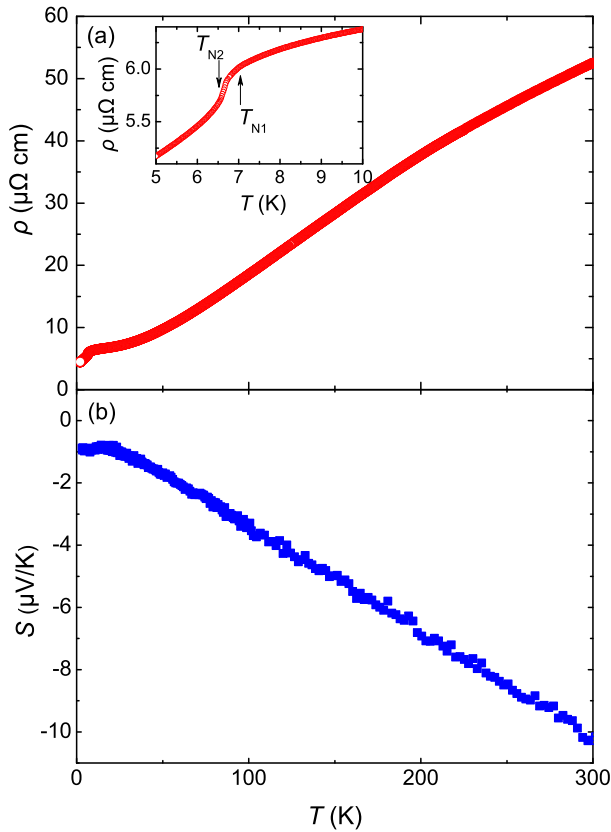}
     \end{center}
     \caption{(Color online) (a) Temperature dependence of the electrical resistivity $\rho(T)$ for single crystals of EuNi$_5$As$_3$. The inset shows  $\rho(T)$ in the vicinity of the magnetic transitions. (b) Temperature dependence of the thermoelectric power for polycrystalline EuNi$_5$As$_3$ in zero applied field.}
     \label{fig4}
\end{figure}

The temperature dependence of the electrical resistivity $\rho(T)$ down to 2~K for single crystals
of EuNi$_5$As$_3$ is shown in Fig.~\ref{fig4}(a) with the current along the a-axis.
At high temperatures, $\rho(T)$ decreases with decreasing temperature, indicating metallic behavior. The residual resistivity at 7~K is $\rho_0$~=~6.01~$\mu\Omega$~cm with a residual resistivity ratio of $RRR\approx8.8$.
 Upon further decreasing the temperature, there is a clear drop in $\rho(T)$ of EuNi$_5$As$_3$ at 6.7~K  due to a magnetic transition, as seen in the magnetic susceptibility $\chi(T)$ and specific heat $C(T)$.
While only a single transition can be clearly resolved in  $\rho(T)$, which is likely $T_{N2}$,  there is also a weaker anomaly at a slightly higher temperature of around 7~K which may correspond to $T_{N1}$.
Figure.~\ref{fig4}(b) shows the temperature dependence of the thermoelectric power $S$ for
polycrystalline EuNi$_5$As$_3$ in the temperature range 2~K-300~K. At room temperature, the
thermoelectric power is about $-10~\mu$V/K, and with decreasing temperature, $S$ linearly
increases before saturating below 20~K, reaching $-1~\mu$V/K$^2$ at 2~K. In the whole temperature range, $S$
remains negative, indicating the dominance of electron-type carriers in EuNi$_5$As$_3$.

\subsection{Field dependence of the magnetic state}

\begin{figure}[t]
     \begin{center}
     \includegraphics[width=3.2in,keepaspectratio]{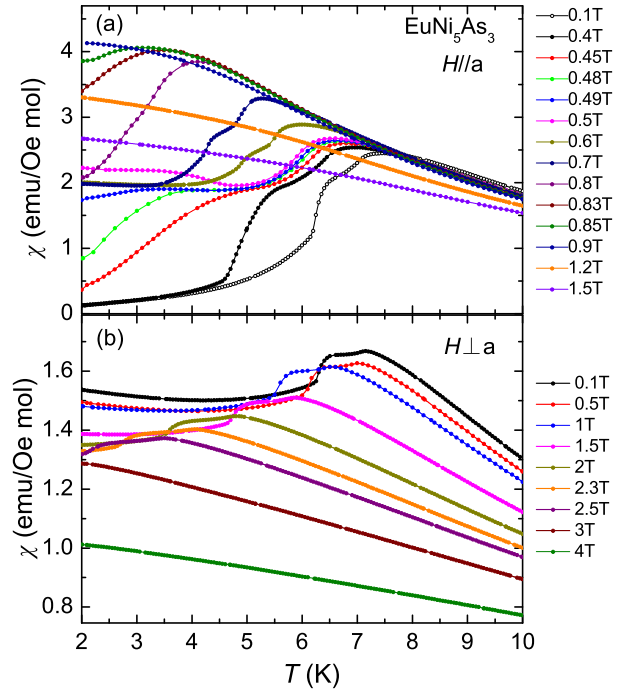}
     \end{center}
     \caption{(Color online)  Temperature dependence of $\chi(T)$ under various applied magnetic fields with (a) $H \|~a$, and (b) $H \bot~a$.}
     \label{fig6}
\end{figure}

The temperature dependence of $\chi(T)$ under various applied magnetic fields is shown in Fig.~\ref{fig6}.
Upon applying magnetic fields, the two magnetic phase transitions $T_{N1}$ and $T_{N2}$ are suppressed to lower temperatures. For $H\parallel a$, $T_{N2}$ becomes broader and weaker at
high fields and is suppressed considerably more rapidly than $T_{N1}$. When 0.5~T is applied the transition corresponding to $T_{N2}$ is no longer observed and instead of a drop   in $\chi(T)$, there is an upturn. This change of behavior may correspond to the emergence of a new magnetic phase and when 0.6~T is applied, two transitions can clearly be seen.
Upon further increasing the field, these two transitions are gradually suppressed to lower temperature before  disappearing
at around 0.9~T. In the case of $H \perp a$, both $T_{N1}$ and $T_{N2}$ are continuously suppressed
and eventually disappear near 3~T.  Therefore the magnetic order is suppressed significantly more rapidly for $H\parallel a$ and  this anisotropy is likely due to the $a$~axis being the easy direction.

\begin{figure}[t]
     \begin{center}
     \includegraphics[width=3.2in,keepaspectratio]{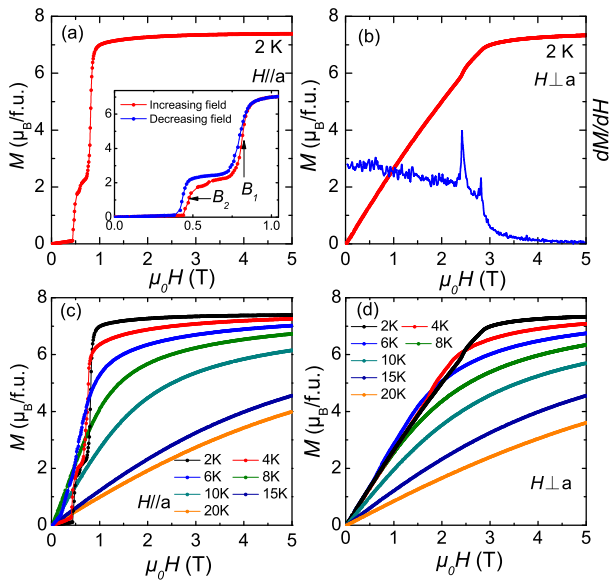}
     \end{center}
     \caption{(Color online) (a) Field dependence of the magnetization $M(H)$ with $H \parallel a$ at 2~K. The inset shows the enlarged low field region measured
     upon both increasing and decreasing the field. The arrows indicate the metamagnetic transitions (b) $M(H)$ and the corresponding derivative $dM/dH$ with $H \bot a$ at 2~K. $M(H)$ at various temperatures is shown for (c) $H \parallel a$ and (d) $H \bot a$.}
     \label{fig7}
\end{figure}

Figure.~\ref{fig7} (a) shows the field dependence of $M(H)$ at 2~K when $H$ is parallel to the $a$ axis.
Below 0.44~T, $M(H)$ increases linearly with magnetic field and there is no hysteresis, which is consistent
with an AFM ground state in EuNi$_5$As$_3$. Upon further increasing the field, $M(H)$ undergoes two sharp
jumps at $B_2~=~0.47$~T and $B_1~=~0.81$~T respectively, consistent with the presence of two metamagnetic transitions.
Hysteresis between field-warming (FW) and field-cooling (FC) can be clearly observed, indicating the first-order nature of these transitions. However upon increasing the temperature towards $T_{N2}$, the hysteresis becomes less pronounced , which indicates that the transitions become more weakly first-order
with increasing temperature. From comparisons with $\chi(T)$ in field, it can be seen that $B_2$ corresponds  to the lower transition observed below 0.5~T and therefore this suggests that  at $B_2$  the magnetic state which appears below $T_{N2}$ is suppressed and there is a transition to the state onsetting at $T_{N1}$. Above the second  transition at $B_1$, the magnetization appears saturated and changes little with increasing field, indicating that this corresponds to a transition from the antiferromagnetically ordered phase to a spin polarized state. 

Figure.~\ref{fig7} (b) shows the field dependence of $M(H)$ at 2~K when $H$ is
perpendicular to the $a$ axis. At low fields below 2.4~T, $M(H)$ shows sub-linear behavior before displaying two anomalies at 2.4~T and 2.8~T, clearly seen as two peaks in $dM/dH$. Unlike the first order transitions observed for
$H \parallel a$, there is no observable hysteresis between the FW and FC measurements.
Figures~\ref{fig7} (c) and (d) show the isothermal magnetization measurements at various temperatures
for two different field orientations. With increasing temperature, all of the metamagnetic transitions are
shifted to lower magnetic fields and broaden upon approaching $T_{N2}$.
At high temperatures above T$_{N1}$, $M(H)$ displays an $S$-shape. Furthermore, the saturated
magnetic moments are $7.38 \mu_B$ and $7.29 \mu_B$ for $H \| \textit{a}$ and $H \bot \textit{a}$, respectively,
very close to the expected Eu$^{2+}$ moment, further indicating that the AFM state arises from the ordering of the Eu$^{2+}$ moments. At 15~K with $H \parallel a$, $M(H)$ is significantly reduced compared to 10~K with a less pronounced $S$-shape, while there is a much smaller change between 15~K and 20~K. This is consistent with the emergence of magnetic fluctuations or short range order upon approaching $T_{N1}$ from higher temperatures, as suggested from specific heat measurements.

\begin{figure}[t]
     \begin{center}
     \includegraphics[width=3.2in,keepaspectratio]{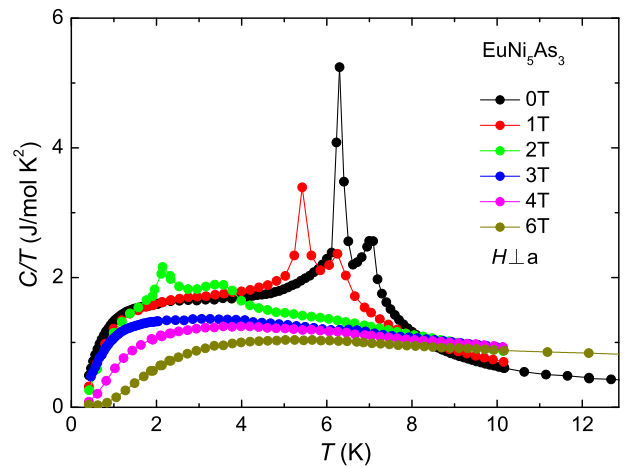}
     \end{center}
     \caption{(Color online) $C/T$ as a function of temperature down to 0.4~K under various magnetic fields which are applied perpendicular to the $a$ axis.}
     \label{fig8}
\end{figure}

Specific heat measurements under applied magnetic fields with $H\perp a$ are shown in Fig.~\ref{fig8}.
 Upon increasing the applied field, the anomalies at $T_{N1}$ and $T_{N2}$ become less pronounced and are
gradually suppressed to lower temperatures, with no transition being observed down to 0.4~K at 3~T.
In addition, the low temperature plateau likely due to the Zeeman splitting of the ground state multiplet is shifted to higher temperatures with increasing field.

\begin{figure}[t]
     \begin{center}
     \includegraphics[width=3.2in,keepaspectratio]{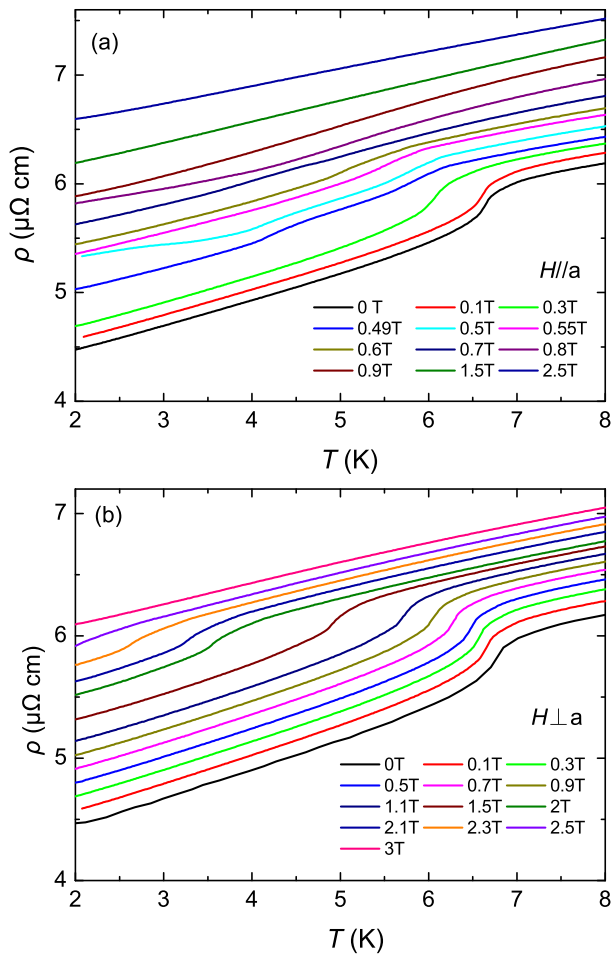}
     \end{center}
     \caption{(Color online)
      Temperature dependence of electrical resistivity at various magnetic fields applied
      (a) parallel to the $a$ axis, and (b) perpendicular to the $a$ axis. The curves are vertically shifted for clarity.}
     \label{fig9}
\end{figure}

Figures~\ref{fig9} (a) and (b) show $\rho(T)$ measured
in various fields with $H \| a$ and $H \bot a$ respectively. For $H \bot a$, below 1.1~T the AFM transition is
slowly suppressed to lower temperature with increasing field before the transition broadens and is rapidly suppressed
in higher fields. No transitions are observed down to 2~K  at
3~T and the resistivity begins to show $\sim T^2$ behavior, as expected for a Fermi liquid. In contrast for $H \| a$, below 0.49~T a sharp drop due to a magnetic transition is observed, along with a weaker anomaly at a slightly higher temperature,  which becomes more pronounced with increasing field. In the vicinity of 0.5~T, two transitions can still be observed but there is now a smaller anomaly at the  lower transition. At higher fields only one transition is clearly seen, which is suppressed to below 2~K upon the application of 0.9~T. From a comparison with zero field specific heat measurements and the magnetic susceptibility, the stronger transition at low fields corresponds to  $T_{N2}$, while the weaker anomaly at higher temperatures is likely $T_{N1}$. When 0.5~T is applied, the lower transition agrees well with the possible new field-induced magnetic phase suggested to emerge in this field region from  magnetic susceptibility measurements [Fig.~\ref{fig6}(a)], after the disappearance of $T_{N2}$.

\begin{figure}[t]
     \begin{center}
     \includegraphics[width=2.5in,keepaspectratio]{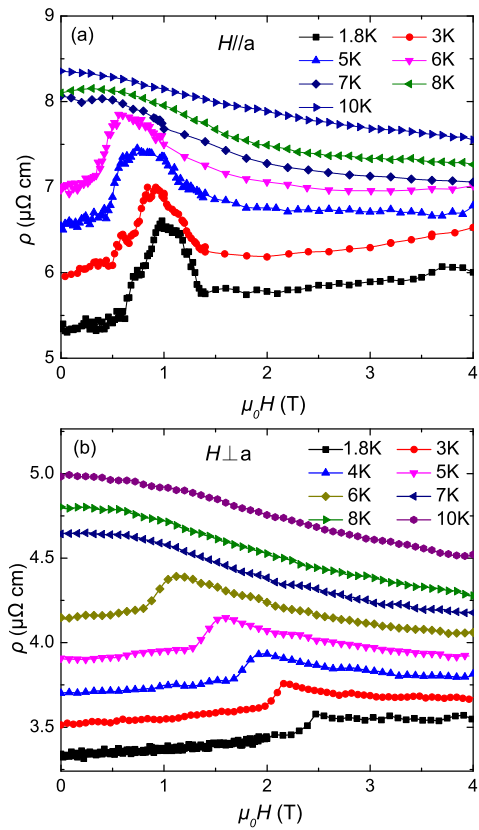}
     \end{center}
     \caption{(Color online) Magnetic field dependence of $\rho(H)$ at different temperatures for  (a) $H \| \textit{a}$ and (b) $H \bot \textit{a}$.}
     \label{fig11}
\end{figure}

The magnetic field dependence of the electrical resistivity $\rho(H)$ at different temperatures is shown
in Fig.~\ref{fig11} (a)($H \| a$) and Fig.~\ref{fig11} (b)($H \bot a$). For $H \bot a$, at 1.8~K $\rho(H)$ has a pronounced
jump at 2.4~T, which also corresponds to the metamagnetic transition seen in  $M(H)$.
With increasing temperature, this jump in $\rho(H)$ decreases to lower fields, reaching 0.9~T at 6~K.
At 7~K near $T_{N1}$, the metamagnetic transition disappears and $\rho(H)$ displays a negative magnetoresistance.
For $H \| a$ at 1.8~K, $\rho(H)$ displays a peak at around 0.9~T.
Both transitions can be attributed to the field-induced metamagnetic transitions seen in the magnetization measurements.
Similar to the $H \bot a$ case, the main peak is shifted to lower fields with increasing temperature.
In the paramagnetic state, a clear negative magnetoresistance is observed due to the reduction of spin
disorder scattering as a result of the alignment of the spins along the applied magnetic field.

\subsection{Eu valence}

\begin{figure}[tb]
     \begin{center}
     \includegraphics[width=2.5in,keepaspectratio]{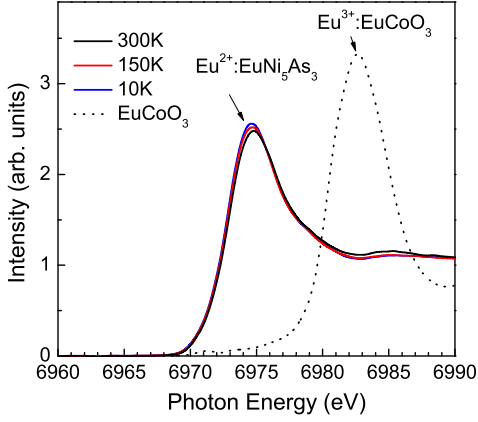}
     \end{center}
     \caption{(Color online) Eu L$_{III}$ PFY-XAS spectra of EuNi$_5$As$_3$ at three different temperatures. The dotted line shows the Eu reference spectrum for EuCoO$_3$(Eu$^{3+}$) for comparison.}
     \label{fig12}
\end{figure}

The value of the effective Eu moment obtained from fitting the magnetic susceptibility,
indicates the localized nature of the Eu in EuNi$_5$As$_3$. To further investigate the Eu valence
of EuNi$_5$As$_3$, we performed Eu L$_{III}$ edge PFY-XAS measurements. In  Fig.~\ref{fig12},  Eu L$_{III}$ spectra are shown at three temperatures, along with the spectrum of EuCoO$_3$ for comparison. The  EuNi$_5$As$_3$ measurements all show a prominent peak at around 6974.7~eV, which is  ascribed to the $2p_{3/2}$ $\rightarrow$ 5d transition in Eu$^{2+}$, agreeing very well with the peak position around 6975~eV observed in divalent EuF$_2$ \cite{JPCMBauer}. If there were a significant Eu$^{3+}$ component, another peak is expected at higher energies, as shown for example by the spectrum of EuCoO$_3$~\cite{HuPRL}, where there is a peak around 6982.6~eV. Since such a prominent peak is not clearly observed in  EuNi$_5$As$_3$, these results indicate that the Eu does not have significantly mixed valence character, but that the valence is very close to +2 at all temperatures down to at least 10~K. This situation is very similar to other divalent magnetically ordered Eu-based compounds, such as,
Eu$T_4$Sb$_{12}$ ($T$=Fe, Ru, Os)~\cite{JPCMBauer,GrytsivPRB}.

\section{Discussion and Summary}

\begin{figure}[tb]
     \begin{center}
     \includegraphics[width=3.2in,keepaspectratio]{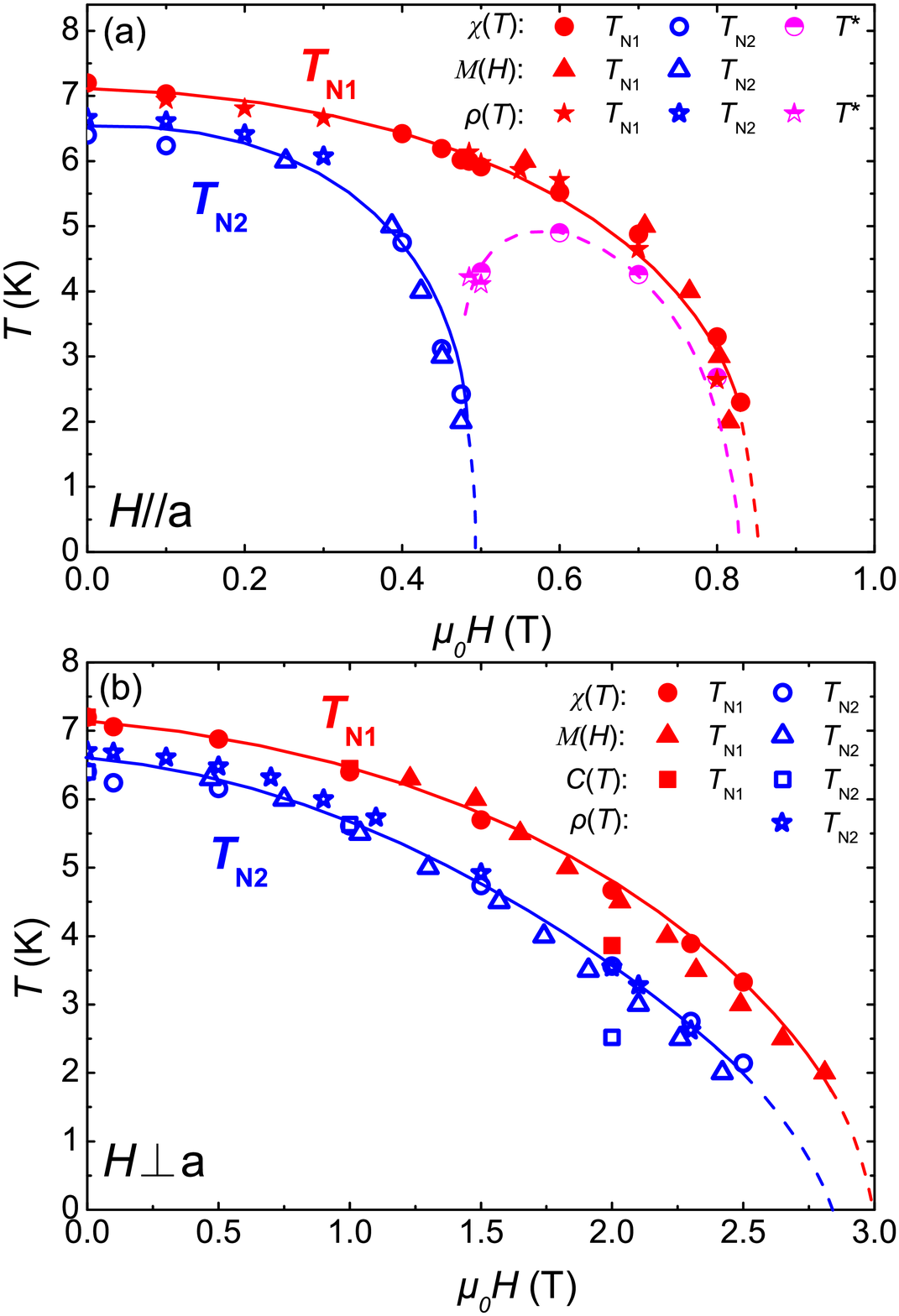}
     \end{center}
     \caption{(Color online) The temperature-field phase diagram of EuNi$_5$As$_3$ for magnetic fields applied (a) parallel to the $a$ axis
     and (b) perpendicular to the $a$ axis. The transitions $T_{N1}$ and $T_{N2}$ are represented
     by red filled and blue open symbols respectively,  while different symbols denote values obtained by different techniques. $T^*$ denotes the possible antiferromagnetic transition induced upon applying magnetic fields parallel to  the $a$ axis.}
     \label{fig13}
\end{figure}

The $T-H$ phase diagram constructed from measurements of the electrical resistivity, magnetic susceptibility and
 specific heat  of EuNi$_5$As$_3$ is shown in Fig.~\ref{fig13} for fields applied parallel and perpendicular to the $a$~axis. The phase boundaries deduced from different measurements are all consistent. As well as the two zero field magnetic transitions at $T_{N1}$ and $T_{N2}$, the  possible new field induced AFM phase above 0.5~T is denoted by $T^*$, which occurs  after the disappearance of $T_{N2}$ for  $H \parallel a$. The temperature evolution of the metamagnetic transitions in $M(H)$  is also shown in the phase diagram, obtained from the maximum of the derivative and the close agreement with the  $\chi(T)$ measurements indicate that these correspond to the suppression of the two AFM phases. In  $\rho(T)$ at zero field and with $H \bot a$, only the position of the  transition  corresponding  to $T_{N2}$ can be clearly resolved from the derivative. For $H \parallel a$ the most pronounced  transition below 0.5~T corresponds to $T_{N2}$, but $T_{N1}$ can still be resolved once a field is applied, up to its suppression at around 0.9~T.

The behavior of EuNi$_5$As$_3$ at ambient pressure appears to be similar to  other Eu based magnetically ordered materials, with a stable Eu$^{2+}$ configuration and a lack of quantum criticality. This is strikingly different from many Ce and
Yb-based Kondo systems, which are well understood on the basis of the Doniach model \cite{Doniach}.  These systems can often be continuously tuned to a quantum critical point using pressure, doping or magnetic fields, where pronounced non-Fermi liquid behavior is observed and there is generally a gradual change of the valence upon increasing the hybridization \cite{Ye,Review1,Review2}.

For Eu systems, the atomic size of the non-magnetic trivalent Eu$^{3+}$ is smaller than that of magnetic divalent Eu$^{2+}$. Applying pressure may destabilize the magnetic Eu$^{2+}$ leading to an abrupt change from Eu$^{2+}$ to Eu$^{3-\delta}$ \cite{SegrePRL}. There are several examples where upon increasing the pressure,  the magnetic phase suddenly disappears at a first-order transition $P_c$, above which the system is non-magnetic with a mixed Eu valence. These features have often been seen in Eu intermetallics, such as EuNi$_2$(Si$_x$Ge$_{1-x}$)$_2$~\cite{Wada1999} and EuRh$_2$Si$_2$ \cite{JPCM2011,Mitsuda2012}.

The antiferromagnetic transition temperatures of EuNi$_5$As$_3$ are 7.2~K and 6.4~K, slightly smaller
than AFM transition temperature of 7.5~K in the isostructural compound EuNi$_5$P$_3$, which has smaller lattice
parameters and therefore corresponds to a positive chemical pressure \cite{Badding_PRB,Fisher1995}.  This indicates that the magnetic phase of EuNi$_5$As$_3$ is likely to be quite robust against pressure.
The antiferromagnetism in EuNi$_5$As$_3$ should be deep inside the AFM region, far away from
the critical line near $P_c$. Therefore, tuning EuNi$_5$P$_3$ with pressure may allow for the system to
either reach a mixed valence state, or possibly even display heavy fermion behavior.

To summarize, we have successfully synthesized single crystalline and polycrystalline EuNi$_5$As$_3$ and
performed a detailed investigation of its crystal structure, physical properties and Eu valence.
From our measurements, EuNi$_5$As$_3$ is an AFM compound with $T_{N1}~=~7.2$~K and a
subsequent AFM transition at $T_{N2}~=~6.4$~K. The
AFM state is sensitive to an applied magnetic field and shows an anisotropic response.
For $H \| a$ , the AFM transitions are all absent at about 0.9~T, while
for $H \bot a$, they are relatively more robust in field than along
the chain direction and are suppressed at about 3~T. Meanwhile both magnetic susceptibility and PFY-XAS measurements indicate that the Eu are strongly divalent with an almost temperature independent valence and there is a lack of evidence for heavy fermion behavior.  To determine the magnetic structure in the ordered phases, neutron scattering measurements are desirable. Furthermore, it may be possible to tune EuNi$_5$As$_3$ or EuNi$_5$P$_3$  towards a valence transition by doping or hydrostatic pressure.

\begin{acknowledgments}
We thank Z. Hu for valuable discussions. This work was supported by the National Natural Science Foundation of China (No. U1632275), National Key Research and Development Program of China (No. 2016YFA0300202), and the Science Challenge Project of China.
\end{acknowledgments}

\end{document}